\documentclass[conference,10pt]{IEEEtran}
\IEEEoverridecommandlockouts
\usepackage{cite}
\usepackage{amsmath,amssymb,amsfonts}
\usepackage{algorithmic}
\usepackage{graphicx}
\usepackage{textcomp}
\usepackage{etoolbox}
\usepackage{float}

\floatstyle{ruled}
\newfloat{algorithm}{tbp}{loa}
\providecommand{\algorithmname}{Algorithm}
\floatname{algorithm}{\protect\algorithmname}

\def\BibTeX{{\rm B\kern-.05em{\sc i\kern-.025em b}\kern-.08em
    T\kern-.1667em\lower.7ex\hbox{E}\kern-.125emX}}

    \makeatletter
\patchcmd{\@maketitle}
  {\addvspace{0.5\baselineskip}\egroup}
  {\addvspace{-1\baselineskip}\egroup}
  {}
  {}

\makeatother

\begin{document}

\title{Optimizing Pilots and Analog Processing for Channel Estimation in
Cell-Free Massive MIMO With One-Bit ADCs}

\author{\IEEEauthorblockN{Seok-Hwan Park} \IEEEauthorblockA{\textit{Dept. of Elect. Engineering} \\
 \textit{Chonbuk National University}\\
 Jeonju, Korea \\
 seokhwan@jbnu.ac.kr} \and \IEEEauthorblockN{Osvaldo Simeone} \IEEEauthorblockA{\textit{Centre for Telecomm. Research} \\
 \textit{King's College London}\\
 London, UK \\
 osvaldo.simeone@kcl.ac.uk} \and \IEEEauthorblockN{Yonina C. Eldar} \IEEEauthorblockA{\textit{Dept. of Elect. Engineering} \\
 \textit{Technion}\\
 Haifa, Israel \\
yonina@ee.technion.ac.il} \and \IEEEauthorblockN{Elza Erkip} \IEEEauthorblockA{\textit{Dept. of ECE}\\
 \textit{New York University}\\
 New York, United States \\
elza@nyu.edu} }
\maketitle
\begin{abstract}
In a cell-free cloud radio access network (C-RAN) architecture, uplink
channel estimation is carried out by a centralized baseband processing
unit (BBU) connected to distributed remote radio heads (RRHs). When
the RRHs have multiple antennas and limited radio front-end resources,
the design of uplink channel estimation is faced with the challenges
posed by reduced radio frequency (RF) chains and one-bit analog-to-digital
converters (ADCs) at the RRHs. This work tackles the problem of jointly
optimizing the pilot sequences and the pre-RF chains analog combiners
with the goal of minimizing the sum of mean squared errors (MSEs)
of the estimated channel vectors at the BBU. The problem formulation
models the impact of the ADC operation by leveraging Bussgang's theorem.
An efficient solution is developed by means of an iterative alternating
optimization algorithm. Numerical results validate the advantages
of the proposed joint design compared to baseline schemes that randomly
choose either pilots or analog combiners.\end{abstract}


\section{Introduction\label{sec:Introduction}}

\let\thefootnote\relax\footnotetext{S.-H. Park was supported by the NRF Korea funded by the Ministry of Science, ICT $\&$ Future Planning through grant 2015R1C1A1A01051825. The work of O. Simeone was partially supported by U.S. NSF through grant 1525629. O. Simeone has also received funding from the European Research Council (ERC) under the European Union's Horizon 2020 research and innovation programme (grant agreement No 725731). Y. C. Eldar received funding from the European Union's Horizon 2020 research and innovation program under grant agreement No. 646804-ERC-COG-BNYQ, and from the Israel Science Foundation under Grant. No. 335/14. E. Erkip was supported by U.S. NSF through grants 1302336 and 1547332.}

In a cell-free cloud radio access network (C-RAN) system, a number
of remote radio heads (RRHs) are deployed to collectively serve users
in the covered area. The RRHs are connected to a baseband processing
unit (BBU) that carries out centralized baseband signal processing
\cite{Quek-et-al}. In a typical 5G deployment, due to the use of
wideband spectrum and massive antenna arrays, it is generally impractical
to equip the RRHs with high-precision analog-to-digital converters
(ADCs) and with one radio frequency (RF) chain per antenna element
due to high cost and power consumption \cite{Sohrabi-Yu}-\!\!\cite{Lee-et-al:TCOM}.
Therefore, RRHs typically have a limited number of RF chains with
limited resolution ADCs. A well-known solution to the problem of limited RF chains is to deploy
a hybrid beamforming architecture, whereby analog combining is applied
prior to ADC operations \cite{Sohrabi-Yu}-\!\!\cite{Ioushua-Eldar}.

A key task in massive MIMO systems is acquiring channel state information
(CSI) at the BBU. This is typically done via uplink training by leveraging
channel reciprocity in Time Division Duplex (TDD) systems. With a
cell-less architecture and centralized processing, the presence of
a large number of users in the covered area implies that the number
of resources allocated for training may not be sufficient to allocate
orthogonal pilot sequences to all users.

In this work, we study channel estimation for a cell-free C-RAN uplink. Following \cite{Ioushua-Eldar}, specifically, we tackle
the problem of jointly optimizing the pilot sequences and the distributed analog
combiners at the RRHs with the goal of minimizing the sum of mean
squared errors (MSEs) of the estimated channel vectors at the BBU.
The problem formulation models the impact of the ADC operation by
leveraging Bussgang's theorem \cite{Bussgang}. We develop an efficient
solution by means of an iterative alternating optimization algorithm.
Numerical results validate the advantages of the proposed
joint design compared to baseline schemes that randomly choose either
pilots or analog combiners.

\textit{Related works:} In \cite{Jacobsson-et-al}\!\!\cite{Li-et-al},
the uplink channel estimation problem was studied for a single-cell
uplink system with low-resolution ADCs and fully-digital, instead
of hybrid, beamforming. The problem of channel estimation for the multi-cell
uplink of massive MIMO systems in the presence of pilot contamination was tackled in \cite{Ioushua-Eldar} under the assumptions that the uplink channel is noiseless, the RRHs use high-resolution ADCs, and they do not cooperate with each other. In \cite{Kang-et-al}, the
design of joint signal and CSI compression for fronthaul transmission was studied for
a C-RAN uplink with finite-capacity fronthaul links under the ergodic
fading channel model. The work \cite{JPark-et-al} studied the optimization of uplink reception with mixed-ADC front-end under the assumption of perfect CSI.

The rest of the paper is organized as follows. The system model for
uplink channel estimation in a cell-free C-RAN system is described in Sec.
\ref{sec:System-Model}. We discuss the problems of jointly optimizing
the pilots and analog processing for channel estimation first under the assumption that the RRHs use high-resolution ADCs in Sec. \ref{sec:optimization-high-resol-ADC} and then with one-bit ADCs in Sec. \ref{sec:Problem-Definition-and}.
In Sec. \ref{sec:Numerical-Results}, we provide numerical results
that validate the advantages of the proposed joint design, and we
conclude the paper in Sec. \ref{sec:Conclusion}.

\textit{Notations:} We denote the circularly symmetric complex Gaussian
distribution with mean $\mbox{\boldmath${\mu}$}$ and covariance matrix
$\mathbf{R}$ as $\mathcal{CN}(\mbox{\boldmath${\mu}$},\bold{R})$.
The set of all $M\times N$ complex matrices is denoted as $\mathbb{C}^{M\times N}$, and $\mathtt{E}(\cdot)$ represents the expectation operator.
We denote the transpose, Hermitian transpose and vectorization operations as
$(\cdot)^{T}$, $(\cdot)^{H}$ and $\mathrm{vec}(\cdot)$, respectively,
and $\mathbf{A}\otimes\mathbf{B}$ represents the Kronecker product
of matrices $\mathbf{A}$ and $\mathbf{B}$. We denote by $\mathbf{I}_{N}$
an $N$-dimensional identity matrix.

\section{System Model\label{sec:System-Model}}

In this section, we describe the system model under study. We consider
the uplink of a cell-free C-RAN system, in which $N_{U}$ single-antenna
user equipments (UEs) communicate with a BBU through $N_{R}$ RRHs. We assume that every
RRH uses $M$ antennas with $L\leq M$ RF chains, each equipped with
a one-bit ADC. Each RRH performs analog combining prior to the ADCs. The fronthaul links connecting the RRHs to BBU are
assumed to have enough capacity to support the transmission of the
ADC outputs. We define the sets $\mathcal{N}_{U}=\{1,\ldots,N_{U}\}$
and $\mathcal{N}_{R}=\{1,\ldots,N_{R}\}$ of UEs' and RRHs' indices.

\subsection{Uplink Channel Model for Pilot Transmission\label{sub:Pilot-Uplink-Channel}}

For uplink channel estimation, each UE $k$ sends a pilot sequence
$\mathbf{s}_{k}=[s_{k,1}\,\cdots\,s_{k,\tau}]^{T}$ during $\tau$
symbols. We impose per-UE transmit power constraints as
\begin{equation}
\frac{1}{\tau}\mathbf{s}_{k}^{H}\mathbf{s}_{k}\leq P_{k},\,\,\mathrm{for}\,\,k\in\mathcal{N}_{U}.\label{eq:per-UE-power-constraints}
\end{equation}
 Assuming a flat-fading channel model, the signal $\mathbf{Y}_{i}\in\mathbb{C}^{M\times\tau}$
received by RRH $i$ can be modeled as
\begin{equation}
\mathbf{Y}_{i}=\sum\nolimits_{k\in\mathcal{N}_{U}}\mathbf{h}_{i,k}\mathbf{s}_{k}^{T}+\mathbf{Z}_{i},\label{eq:received-signal-RRH-i}
\end{equation}
where $\mathbf{h}_{i,k}\in\mathbb{C}^{M\times1}$ denotes the channel
vector from UE $k$ to RRH $i$, and $\mathbf{Z}_{i}$ represents
the additive noise matrix distributed as $\mathbf{z}_{i}=\mathrm{vec}(\mathbf{Z}_{i})\sim\mathcal{CN}(\mathbf{0},\sigma_{i}^{2}\mathbf{I}_{M\tau})$.
As in \cite{Ioushua-Eldar}, we model each channel vector $\mathbf{h}_{i,k}$ as
\begin{equation}
\mathbf{h}_{i,k}=\sqrt{\rho_{i,k}}\,\mathbf{Q}_{i}^{1/2}\mathbf{h}_{i,k}^{w},\label{eq:channel-correlation-model}
\end{equation}
where $\rho_{i,k}=1/(1+(D_{i,k}/10)^3)$ denotes the pathloss, with
$D_{i,k}$ being the distance between RRH $i$ and UE $k$, $\mathbf{Q}_{i}$
represents the receive correlation matrix of RRH $i$,
and $\mathbf{h}_{i,k}^{w}$ is a spatially white channel vector distributed
as $\mathbf{h}_{i,k}^{w}\sim\mathcal{CN}(\mathbf{0},\mathbf{I}_{M})$.
We assume that the channel vectors $\mathbf{h}_{i,k}$ are independent
across the indices $i$ and $k$. The discussion can be generalized
to the case where the channel vectors from different UEs are correlated
\cite{Ioushua-Eldar}.

\subsection{Reduced RF Chain and Analog Combining\label{sub:Reduced-RF-Chain}}

Since each RRH $i$ uses only $L$ RF chains, analog combining is carried out at RRH $i$
via a matrix $\mathbf{W}_{i}\in\mathbb{C}^{L\times M}$. Analog combining
maps the $M$ received signals into an $L$-dimensional vector
\begin{equation}
\tilde{\mathbf{Y}}_{i}=\mathbf{W}_{i}\mathbf{Y}_{i},\label{eq:analog-combining}
\end{equation}
with $\tilde{\mathbf{Y}}_{i}\in\mathbb{C}^{L\times\tau}$. The condition
on the analog combining matrix $\mathbf{W}_{i}$ depends on the specific
architecture of the analog network \cite{Ioushu-Eldar:SU}$\!\!$\cite{Ioushua-Eldar}. In this
work, we consider fully-connected phase shifters network  so that
the matrix $\mathbf{W}_{i}$ is subject to constant modulus constraints
stated as \cite{Sohrabi-Yu}
\begin{equation}
\left|\mathbf{W}_{i}(a,b)\right|^{2}=1,\,\,\mathrm{for}\,\,a\in\mathcal{L},\,b\in\mathcal{M},\label{eq:constant-modulus-constraints}
\end{equation}
where $\mathbf{W}_{i}(a,b)$ indicates the $(a,b)$th element of $\mathbf{W}_{i}$,
and $\mathcal{M}=\{1,\ldots,M\}$ and $\mathcal{L}=\{1,\ldots,L\}$
denote the sets of antennas' and RF chains' indices, respectively.

For mathematical convenience, we also introduce the vectorized version
$\tilde{\mathbf{y}}_{i}\in\mathbb{C}^{L\tau\times1}$ of the signal
$\tilde{\mathbf{Y}}_{i}$ as
\begin{equation}
\tilde{\mathbf{y}}_{i}=\mathrm{vec}(\tilde{\mathbf{Y}}_{i})=\sum\nolimits_{k\in\mathcal{N}_{U}}\mathbf{B}_{k,i}\mathbf{h}_{i,k}+\tilde{\mathbf{z}}_{i},\label{eq:received-signal-vectorized}
\end{equation}
where we defined the notations $\mathbf{B}_{k,i}=\mathbf{s}_{k}\otimes\mathbf{W}_{i}$
and $\tilde{\mathbf{z}}_{i}=\mathrm{vec}(\mathbf{W}_{i}\mathbf{Z}_{i})=(\mathbf{I}_{\tau}\otimes\mathbf{W}_{i})\mathrm{vec}(\mathbf{Z}_{i})$.
Here the effective noise vector $\tilde{\mathbf{z}}_{i}$ is distributed
as $\tilde{\mathbf{z}}_{i}\sim\mathcal{CN}(\mathbf{0},\mathbf{C}_{\tilde{\mathbf{z}}_{i}})$
with $\mathbf{C}_{\tilde{\mathbf{z}}_{i}}=\sigma_{i}^{2}(\mathbf{I}_{\tau}\otimes\mathbf{W}_{i})(\mathbf{I}_{\tau}\otimes\mathbf{W}_{i})^{H}$.
We note that the covariance $\mathbf{C}_{\tilde{\mathbf{y}}_{i}}=\mathtt{E}[\tilde{\mathbf{y}}_{i}\tilde{\mathbf{y}}_{i}^{H}]$
of vector $\tilde{\mathbf{y}}_{i}$ is
\begin{equation}
\mathbf{C}_{\tilde{\mathbf{y}}_{i}}=\sum\nolimits_{k\in\mathcal{N}_{U}}\rho_{i,k}\mathbf{B}_{k,i}\mathbf{Q}_{i}\mathbf{B}_{k,i}^{H}+\mathbf{C}_{\tilde{\mathbf{z}}_{i}}.\label{eq:covariance-received-signal}
\end{equation}

\subsection{One-Bit ADC\label{sub:One-Bit-ADC}}

Each RRH $i$ quantizes the in-phase and quadrature (IQ) components
of the elements of the vector $\tilde{\mathbf{y}}_{i}$ using one-bit
ADCs. As in \cite{Jacobsson-et-al}\!\!\cite{Li-et-al}, we model the
impact of one-bit ADC using Bussgang's theorem \cite{Bussgang}. Accordingly,
the ADC output vector, denoted by $\hat{\mathbf{y}}_{i}$, is statistically
equivalent to
\begin{equation}
\hat{\mathbf{y}}_{i}=\mathbf{A}_{i}\tilde{\mathbf{y}}_{i}+\mathbf{q}_{i},\label{eq:Bussgang-decomposition}
\end{equation}
where the transformation matrix $\mathbf{A}_{i}$ is equal to
\begin{equation}
\mathbf{A}_{i}=\sqrt{\frac{1}{2}} \mathbf{\Sigma}_{\tilde{\mathbf{y}}_{i}}^{-1/2},\label{eq:transform-matrix-Bussgang}
\end{equation}
and vector $\mathbf{q}_{i}$ represents the quantization noise uncorrelated
to the input signal $\tilde{\mathbf{y}}_{i}$. The matrix $\mathbf{\Sigma}_{\tilde{\mathbf{y}}_{i}}$ denotes
a diagonal matrix that contains only the diagonal elements of $\mathbf{C}_{\tilde{\mathbf{y}}_{i}}$.
Furthermore, the covariance matrix $\mathbf{C}_{\mathbf{q}_{i}}=\mathtt{E}[\mathbf{q}_{i}\mathbf{q}_{i}^{H}]$
of vector $\mathbf{q}_{i}$ is equal to
\begin{align}
\mathbf{C}_{\mathbf{q}_{i}} & =\mathbf{C}_{\hat{\mathbf{y}}_{i}}-\mathbf{A}_{i}\mathbf{C}_{\tilde{\mathbf{y}}_{i}}\mathbf{A}_{i}^{H},\label{eq:covariance-quantization-noise}
\end{align}
with the covariance matrix $\mathbf{C}_{\hat{\mathbf{y}}_{i}}=\mathtt{E}[\hat{\mathbf{y}}_{i}\hat{\mathbf{y}}_{i}^{H}]$
given by
\begin{equation}
\mathbf{C}_{\hat{\mathbf{y}}_{i}}=\frac{2}{\pi}\left[\begin{array}{c}
\arcsin\left(\mathbf{\Sigma}_{\tilde{\mathbf{y}}_{i}}^{-1/2}\Re\left\{ \mathbf{C}_{\tilde{\mathbf{y}}_{i}}\right\} \mathbf{\Sigma}_{\tilde{\mathbf{y}}_{i}}^{-1/2}\right)+\\
j\arcsin\left(\mathbf{\Sigma}_{\tilde{\mathbf{y}}_{i}}^{-1/2}\Im\left\{ \mathbf{C}_{\tilde{\mathbf{y}}_{i}}\right\} \mathbf{\Sigma}_{\tilde{\mathbf{y}}_{i}}^{-1/2}\right)
\end{array}\right].\label{eq:covariance-ADC-output}
\end{equation}
We note that the matrices $\mathbf{A}_{i}$ and $\mathbf{C}_{\mathbf{q}_{i}}$
depend both on the pilots $\mathbf{S}=\{\mathbf{s}_{k}\}_{k\in\mathcal{N}_{U}}$
and the analog combining matrix $\mathbf{W}_{i}$, since the covariance
matrix $\mathbf{C}_{\tilde{\mathbf{y}}_{i}}$ defined in (\ref{eq:covariance-received-signal})
is a function of $\{\mathbf{B}_{k,i}\}_{k\in\mathcal{N}_{U}}$ with
$\mathbf{B}_{k,i}=\mathbf{s}_{k}\otimes\mathbf{W}_{i}$.

\subsection{Channel Estimation\label{sub:LMMSE-Channel-Estimation}}

The BBU estimates all the channel vectors $\{\mathbf{h}_{i,k}\}_{i\in\mathcal{N}_{R},k\in\mathcal{N}_{U}}$
based on the quantized signals $\hat{\mathbf{y}}=[\hat{\mathbf{y}}_{1}^{H}\,\cdots\,\hat{\mathbf{y}}_{N_{R}}^{H}]^{H}$
collected from the RRHs:
\begin{equation}
\hat{\mathbf{y}}=\sum\nolimits_{k\in\mathcal{N}_{U}}\mathbf{A}\mathbf{B}_{k}\mathbf{h}_{k}+\mathbf{A}\tilde{\mathbf{z}}+\mathbf{q},\label{eq:ADC-collected}
\end{equation}
where we defined the notations $\mathbf{A}=\mathrm{diag}(\mathbf{A}_{1},\ldots,\mathbf{A}_{N_{R}})$,
$\mathbf{B}_{k}=\mathrm{diag}(\mathbf{B}_{k,1},\ldots,\mathbf{B}_{k,N_{R}})$,
$\mathbf{h}_{k}=[\mathbf{h}_{1,k}^{H}\,\cdots\,\mathbf{h}_{N_{R},k}^{H}]^{H}$,
$\tilde{\mathbf{z}}=[\tilde{\mathbf{z}}_{1}^{H}\,\cdots\,\tilde{\mathbf{z}}_{N_{R}}^{H}]^{H}\sim\mathcal{CN}(\mathbf{0},\mathbf{C}_{\tilde{\mathbf{\mathbf{z}}}})$
and $\mathbf{q}=[\mathbf{q}_{1}^{H}\,\cdots\,\mathbf{q}_{N_{R}}^{H}]^{H}\sim\mathcal{CN}(\mathbf{0},\mathbf{C}_{\mathbf{q}})$
with $\mathbf{C}_{\tilde{\mathbf{\mathbf{z}}}}=\mathrm{diag}(\mathbf{C}_{\tilde{\mathbf{\mathbf{z}}}_{1}},\ldots,\mathbf{C}_{\tilde{\mathbf{\mathbf{z}}}_{N_{R}}})$
and $\mathbf{C}_{\mathbf{q}}=\mathrm{diag}(\mathbf{C}_{\mathbf{q}_{1}},\ldots,\mathbf{C}_{\mathbf{q}_{N_{R}}})$.

As in \cite{Ioushua-Eldar}, we assume that the BBU applies a linear
channel estimator to the signal $\hat{\mathbf{y}}$ so that the estimate
$\hat{\mathbf{h}}_{k}$ of $\mathbf{h}_{k}$ is given as
\begin{equation}
\hat{\mathbf{h}}_{k}=\mathbf{F}_{k}\hat{\mathbf{y}},\label{eq:linear-estimate}
\end{equation}
with a linear filter matrix $\mathbf{F}_{k}\in\mathbb{C}^{MN_{R}\times L\tau N_{R}}$.
For given $\mathbf{S}$, $\mathbf{W}=\{\mathbf{W}_{i}\}_{i\in\mathcal{N}_{R}}$
and $\mathbf{F}_{k}$, the MSE $\varepsilon_{k}=\mathtt{E}[||\hat{\mathbf{h}}_{k}-\mathbf{h}_{k}||^{2}]$
is equal to
\begin{align}
\varepsilon_{k} & =e_{k}\left(\mathbf{S},\mathbf{W},\mathbf{F}_{k}\right)\label{eq:MSE-UE-k}\\
= & \mathrm{tr}\left(\left(\mathbf{F}_{k}\mathbf{A}\mathbf{B}_{k}-\mathbf{I}_{MN_{R}}\right)\mathbf{\Theta}_{k}\left(\mathbf{F}_{k}\mathbf{A}\mathbf{B}_{k}-\mathbf{I}_{MN_{R}}\right)^{H}\right)\nonumber \\
 & +\sum\nolimits_{l\in\mathcal{N}_{U}\setminus\{k\}}\mathrm{tr}\left(\mathbf{F}_{k}\mathbf{A}\mathbf{B}_{l}\mathbf{\Theta}_{l}\mathbf{B}_{l}^{H}\mathbf{A}^{H}\mathbf{F}_{k}^{H}\right)\nonumber \\
 & +\mathrm{tr}\left(\mathbf{F}_{k}\mathbf{A}\mathbf{C}_{\tilde{\mathbf{\mathbf{z}}}}\mathbf{A}^{H}\mathbf{F}_{k}^{H}\right)+\mathrm{tr}\left(\mathbf{F}_{k}\mathbf{C}_{\mathbf{q}}\mathbf{F}_{k}^{H}\right),\nonumber
\end{align}
with the definition $\mathbf{\Theta}_{k}=\mathrm{diag}(\{\rho_{i,k}\mathbf{Q}_{i}\}_{i\in\mathcal{N}_{R}})$.
We aim at minimizing the sum-MSE $\varepsilon_{\text{sum}} = \sum_{k\in\mathcal{N}_U} \varepsilon_{k}$ over the pilots $\mathbf{S}$, the analog combiners $\mathbf{W}$ and the digital filter matrices $\mathbf{F}=\{\mathbf{F}_{k}\}_{k\in\mathcal{N}_{U}}$.

\section{Optimization With High-Resolution ADCs} \label{sec:optimization-high-resol-ADC}

In this section, we discuss the joint optimization of the pilots $\mathbf{S}$ and analog processing $\mathbf{W}$ under the assumption that the RRHs use high-resolution ADCs (i.e., $\mathbf{A}_{i}=\mathbf{I}_{L\tau}$, $\mathbf{C}_{\mathbf{q}_{i}}=\mathbf{0}$ and $\hat{\mathbf{y}}_{i}=\tilde{\mathbf{y}}_{i}$ for $i\in\mathcal{N}_R$). Furthermore, as in \cite{Ioushua-Eldar}, we assume that the uplink channel is noise-free, i.e., $\sigma_i^2=0$, $i\in\mathcal{N}_R$.

Define the channel vector $\mathbf{h}_{R,i}=[\mathbf{h}_{i,1}^H \cdots \mathbf{h}_{i,N_U}^H]^H$ for RRH $i$ and whole channel vector $\mathbf{h}_R=[\mathbf{h}_{R,1}^H \cdots \mathbf{h}_{R,N_R}^H]^H$.
Following the same steps as in \cite[Sec. III]{Ioushua-Eldar}, we can write the MMSE estimate of the whole channel vector $\mathbf{h}_R$ as
\begin{align} \label{eq:MMSE-estimate-whole}
\hat{ \mathbf{h} }_{R,\text{MMSE}} = \bar{\mathbf{R}} \bar{\mathbf{B}}^H_R \left( \bar{\mathbf{B}}_R \bar{\mathbf{R}} \bar{\mathbf{B}}^H_R  \right)^{-1} \hat{\mathbf{y}},
\end{align}
where we have defined the notations $\bar{\mathbf{R}} = \text{diag}(\{\mathbf{R}_i\}_{i\in\mathcal{N}_R})$ and $\bar{\mathbf{B}}_R = \text{diag}(\{\mathbf{B}_{R,i}\}_{i\in\mathcal{N}_R})$ with $\mathbf{R}_i = \mathbf{P}_i \otimes \mathbf{Q}_i$, $\mathbf{P}_i = \text{diag}(\{\rho_{i,k}\}_{k\in\mathcal{N}_U})$, $\mathbf{B}_{R,i} = \bar{\mathbf{S}} \otimes \mathbf{W}_i$ and $\bar{\mathbf{S}} = [\mathbf{s}_1 \cdots \mathbf{s}_{N_U}]$.
The estimate in (\ref{eq:MMSE-estimate-whole}) can be decoupled across RRHs, i.e.,
\begin{align} \label{eq:MMSE-estimate-RRH-i}
\hat{ \mathbf{h} }_{R,i, \text{MMSE}} = \mathbf{R}_i \mathbf{B}^H_{R,i} \left( \mathbf{B}_{R,i} \mathbf{R}_i \mathbf{B}^H_{R,i}  \right)^{-1} \hat{\mathbf{y}}_i,
\end{align}
for $i\in\mathcal{N}_R$, due to the independence of the channel vectors $\mathbf{h}_{R,1},\ldots,\mathbf{h}_{R,N_R}$ and distributed analog processing at RRHs.

The sum-MSE $\varepsilon_{\text{sum}}=\sum\nolimits_{i\in\mathcal{N}_R} \mathtt{E} || \mathbf{h}_{R,i} - \hat{ \mathbf{h} }_{R,i, \text{MMSE}} ||^2$ can hence be decomposed as
\begin{align}
&\varepsilon_{\text{sum}} = \sum\nolimits_{i\in\mathcal{N}_R} \big[
\mathrm{tr}\left(\mathbf{R}_i\right)
- \mathrm{tr}\left(  \mathbf{J}_i  \right) \cdot \mathrm{tr}\left( \mathbf{K}_i \right)
\big], \label{eq:MMSE-noiseless}
\end{align}
with matrices $\mathbf{J}_i = \mathbf{W}_i \mathbf{Q}_i^2 \mathbf{W}_i^H ( \mathbf{W}_i\mathbf{Q}_i\mathbf{W}_i )^{-1}$ and
$\mathbf{K}_i = \bar{\mathbf{S}} \mathbf{P}_i^2 \bar{\mathbf{S}}^H (\bar{\mathbf{S}} \mathbf{P}_i \bar{\mathbf{S}}^H)^{-1}$.
Since the covariance matrices $\mathbf{R}_i$ are fixed, the problem of minimizing the sum-MSE in (\ref{eq:MMSE-noiseless}) is equivalent to that of maximizing $\sum_{i\in\mathcal{N}_R} \mathrm{tr}(  \mathbf{J}_i  )\cdot \mathrm{tr}(\mathbf{K}_i) $.

In order to minimize the sum-MSE $\varepsilon_{\text{sum}}$, the analog combiner $\mathbf{W}_i$ of each RRH $i$ can be separately optimized according to the problem:
\begin{align}
\!\!\underset{\mathbf{W}_i}{\mathrm{maximize}} \,\, \mathrm{tr}(  \mathbf{J}_i  ) \,\,\,\,\mathrm{s.t.} \,\, \textrm{(\ref{eq:constant-modulus-constraints}).} \label{eq:optimizing-analog-combiner-noiseless}
\end{align}
The problem (\ref{eq:optimizing-analog-combiner-noiseless}) is the same as that in \cite[Eq. (16)]{Ioushua-Eldar} and hence can be tackled by using the approach proposed in \cite[Sec. IV]{Ioushua-Eldar}.

Given the optimal analog combiners, the optimization over the pilots $\mathbf{S}$ amounts to the maximization of $\sum\nolimits_{i\in\mathcal{N}_R} \!\! w_i \cdot  \mathrm{tr}(\mathbf{K}_i)$, where $w_i=\mathrm{tr}(  \mathbf{J}_i  )$ is now a fixed constant.
To the best of our knowledge, as was also reported in \cite{Ioushua-Eldar}, there is no known solution to this problem except for special cases with $N_R=1$ or $\tau=1$ or $\mathbf{P}_i = \mathbf{P}$ for all $i\in\mathcal{N}_R$. Instead, we propose to adopt the greedy sum of ratio traces maximization (GSRTM) algorithm \cite[Sec. V-C]{Ioushua-Eldar} to find an efficient solution of the problem.


\section{Optimization With One-bit ADCs\label{sec:Problem-Definition-and}}



In this section, we tackle the problem of jointly optimizing the pilots $\mathbf{S}$, the analog combiners
$\mathbf{W}$ and the digital filter matrices $\mathbf{F}$ under the more challenging scenario with one-bit, instead of high-resolution, ADCs. Also, unlike Sec. \ref{sec:optimization-high-resol-ADC}, we assume that the uplink channel is noisy, i.e., $\sigma_i^2 > 0$, $i\in \mathcal{N}_R$.
The problem at hand can be stated as\begin{subequations}\label{eq:problem-original}
\begin{align}
\!\!\underset{\mathbf{S},\mathbf{W},\mathbf{F}}{\mathrm{minimize}} & \sum\nolimits_{k\in\mathcal{N}_{U}}e_{k}\left(\mathbf{S},\mathbf{W},\mathbf{F}_{k}\right)\label{eq:problem-original-cost}\\
\!\!\mathrm{s.t.}\,\,\,\, & \frac{1}{\tau}\mathbf{s}_{k}^{\dagger}\mathbf{s}_{k}\leq P_{k},\,\,\mathrm{for}\,k\in\mathcal{N}_{U},\label{eq:problem-original-power-constraint}\\
 & \left|\mathbf{W}_{i}(a,b)\right|^{2}\!=\!1,\,\mathrm{for}\,a\in\mathcal{L},\,b\in\mathcal{M},\,i\in\mathcal{N}_{R}.\label{eq:problem-original-modulus-constraint}
\end{align}
\end{subequations}
We note that, with the channel noise and quantization distortion, the sum-MSE in (\ref{eq:problem-original-cost}) does not decouple as in (\ref{eq:MMSE-noiseless}) even if we plug the optimal (MMSE) filter $\mathbf{F}$ into (\ref{eq:problem-original-cost}). Therefore, we propose here to solve the problem alternately over the variables $\mathbf{S}$,
$\mathbf{W}$ and $\mathbf{F}$.


\subsection{Proposed Optimization\label{sub:Proposed-Optimization}}

To start, we observe that, if we fix in (\ref{eq:problem-original})
the transformation matrices $\mathbf{A}$ and the covariance matrices
$\mathbf{C}_{\mathbf{q}}$ and relax the constraint (\ref{eq:problem-original-modulus-constraint})
as $\left|\mathbf{W}_{i}(a,b)\right|^{2}\leq1$ for $a\in\mathcal{L}$,
$b\in\mathcal{M}$ and $i\in\mathcal{N}_{R}$, the problem becomes
separately convex with respect to the variables $\mathbf{S}$, $\mathbf{W}$
and $\mathbf{F}$ \cite{Xu-Yin}. This observation motivates us to
derive an alternating optimization algorithm. Note that fixing matrices
$\mathbf{A}$ and $\mathbf{C}_{\mathbf{q}}$ ignores their dependence
on variables $\mathbf{S}$ and $\mathbf{W}$ as in (\ref{eq:transform-matrix-Bussgang})
and (\ref{eq:covariance-quantization-noise}).

The algorithm, which is described in Algorithm 1, solves sequentially
the convex problems obtained from (\ref{eq:problem-original}) by
restricting the optimization variables only to $\mathbf{W}$, $\mathbf{S}$
and $\mathbf{F}$. When solving the convex problems with respect to
$\mathbf{W}$, constraint (\ref{eq:problem-original-modulus-constraint})
is relaxed as $\left|\mathbf{W}_{i}(a,b)\right|^{2}\leq1$ for $a\in\mathcal{L}$,
$b\in\mathcal{M}$ and $i\in\mathcal{N}_{R}$, and the resulting problem can be solved separately for every RRH $i$ without loss of optimality. A feasible solution
is obtained by using the projection approach in \cite[Eq. (18)]{Ioushua-Eldar}.
Also, the optimal linear filter $\mathbf{F}_{k}$, $k\in\mathcal{N}_{U}$,
for fixed other variables is obtained in closed form as
\begin{align}
 & \mathbf{F}_{k,\mathrm{MMSE}}=\mathtt{E}\left[\mathbf{h}_{k}\hat{\mathbf{y}}^{H}\right]\mathtt{E}\left[\hat{\mathbf{y}}\hat{\mathbf{y}}^{H}\right]^{-1}\label{eq:LMMSE-filter}\\
 & =\mathbf{\Theta}_{k}\mathbf{B}_{k}^{H}\mathbf{A}^{H}\left(\sum\nolimits_{l\in\mathcal{N}_{U}}\mathbf{A}\mathbf{B}_{l}\mathbf{\Theta}_{l}\mathbf{B}_{l}^{H}\mathbf{A}^{H}+\mathbf{A}\mathbf{C}_{\tilde{\mathbf{\mathbf{z}}}}\mathbf{A}^{H}+\mathbf{C}_{\mathbf{q}}\right).\nonumber
\end{align}
The step size sequence $\gamma^{t}$ is selected to be decreasing
with the iteration number $t$ as in \cite[Eq. (5)]{Scutari-et-al},
as a means to improve the empirical convergence properties of the
algorithm.

\begin{algorithm}
\caption{Iterative optimization algorithm for problem (\ref{eq:problem-original})}

\textbf{Initialization:}

\textbf{1.} Initialize the pilot sequence $\mathbf{S}^{(1)}$ and
analog combining variables $\mathbf{W}^{(1)}$ such that the conditions
(\ref{eq:problem-original-power-constraint}) and (\ref{eq:problem-original-modulus-constraint})
are satisfied.

\textbf{2.} Initialize the matrices $\mathbf{A}_{i}^{(1)}$ and $\mathbf{C}_{\mathbf{q}_{i}}^{(1)}$,
$i\in\mathcal{N}_{R}$, according to (\ref{eq:transform-matrix-Bussgang})
and (\ref{eq:covariance-quantization-noise}), respectively, for fixed
$\mathbf{S}^{(1)}$ and $\mathbf{W}^{(1)}$, and set $t\leftarrow1$.

\textbf{3.} Initialize the filter matrices $\mathbf{F}_{k}^{(1)}$,
$k\in\mathcal{N}_{U}$, according to (\ref{eq:LMMSE-filter}) for
fixed $\mathbf{S}^{(1)}$, $\mathbf{W}^{(1)}$, $\mathbf{A}^{(1)}$
and $\mathbf{C}_{\mathbf{q}}^{(1)}$.

\textbf{Iteration:}

\textbf{4.} Update the pilot sequences $\mathbf{S}^{(t+1)}$ as $\mathbf{S}^{(t+1)}\leftarrow\mathbf{S}^{(t)}+\gamma^{t}(\mathbf{S}^{\prime}-\mathbf{S}^{(t)})$,
where $\mathbf{S}^{\prime}$ denotes a solution of the problem (\ref{eq:problem-original})
for fixed $\mathbf{W}^{(t)}$, $\mathbf{A}^{(t)}$, $\mathbf{C}_{\mathbf{q}}^{(t)}$
and $\mathbf{F}^{(t)}$.

\textbf{5.} Update the analog combiners $\mathbf{W}^{(t+1)}$ as $\mathbf{W}^{(t+1)}\leftarrow\mathrm{proj}(\mathbf{W}^{(t)}+\gamma^{t}(\mathbf{W}^{\prime}-\mathbf{W}^{(t)}))$,
where $\mathbf{W}^{\prime}$ denotes a solution of the problem (\ref{eq:problem-original})
for fixed $\mathbf{S}^{(t+1)}$, $\mathbf{A}^{(t)}$, $\mathbf{C}_{\mathbf{q}}^{(t)}$
and $\mathbf{F}^{(t)}$, and $\mathrm{proj}(\cdot)$ denotes the projection
onto the space of matrices that satisfy (\ref{eq:problem-original-modulus-constraint}).

\textbf{6.} Update the matrices $\mathbf{A}_{i}^{(t+1)}$ and $\mathbf{C}_{\mathbf{q}_{i}}^{(t+1)}$,
$i\in\mathcal{N}_{R}$, according to (\ref{eq:transform-matrix-Bussgang})
and (\ref{eq:covariance-quantization-noise}), respectively, for fixed
$\mathbf{S}^{(t+1)}$ and $\mathbf{W}^{(t+1)}$.

\textbf{7.} Update the filter matrices $\mathbf{F}_{k}^{(t+1)}$,
$k\in\mathcal{N}_{U}$, according to (\ref{eq:LMMSE-filter}) for
fixed $\mathbf{S}^{(t+1)}$, $\mathbf{W}^{(t+1)}$, $\mathbf{A}^{(t+1)}$
and $\mathbf{C}_{\mathbf{q}}^{(t+1)}$.

\textbf{8.} Stop if a convergence criterion is satisfied. Otherwise,
set $t\leftarrow t+1$ and go back to Step 4.
\end{algorithm}


\section{Numerical Results\label{sec:Numerical-Results}}

In this section, we provide numerical results that validate the effectiveness
of the proposed joint design of the pilot sequences and analog combining
matrices for the uplink of cell-free C-RAN with one-bit ADCs. For
performance evaluation, we assume that $N_{U}$ UEs and $N_{R}$ RRHs
are uniformly distributed within a square area of the side length
equal to 100 m. As in \cite{Shiu-et-al}\!\!\cite{Ioushua-Eldar:SPAWC},
the correlation matrix $\mathbf{R}_{i,k}$ in (\ref{eq:channel-correlation-model})
is given as $\mathbf{R}_{i,k}(a,b)=J_{0}(2\pi\left|a-b\right|\sin(d_{i}/\lambda_{i})/\Delta_{i}),$
where $J_{0}(\cdot)$ denotes the zero-th order Bessel function, and
we set $d_{i}/\lambda_{i}=0.5$ and $\Delta_{i}=25$ \cite{Ioushua-Eldar:SPAWC}.

For comparison, we consider the performance of the following reference
schemes: \textit{i)} \textit{Fully random:} Pilot sequences $\mathbf{S}$
and analog combining matrices $\mathbf{W}$ are randomly chosen; \textit{ii)
Optimized analog combining with random pilots:} Analog combining matrices
$\mathbf{W}$ are optimized for randomly selected pilot sequences
$\mathbf{S}$; \textit{iii) Optimized pilots with random analog combining:}
Pilot signals $\mathbf{S}$ are optimized for randomly selected analog
combining matrices $\mathbf{W}$; and \textit{iv)} \textit{Proposed
joint design:} Pilot sequences $\mathbf{S}$ and analog combining
matrices $\mathbf{W}$ are jointly optimized.

The algorithm proposed in Sec. \ref{sub:Proposed-Optimization} is
used for the last scheme, while the other reference approaches are
implemented adding the indicated linear constraints to the optimization
problem (\ref{eq:problem-original}).

\begin{figure}
\centering\includegraphics[width=8.65cm,height=6.15cm]{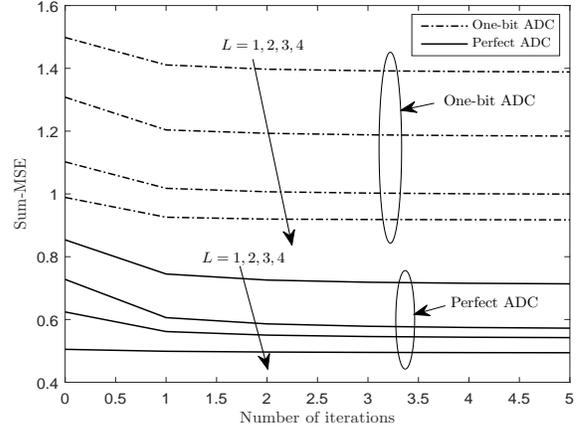}\caption{{\scriptsize{}{}\label{fig:graph-iterations}}{\footnotesize{}{}Average
sum-MSE versus the number of iterations ($N_{U}=6$, $N_{R}=2$, $M=4$,
$\tau=2$ and 10 dB SNR).}}
\end{figure}

We first observe in Fig. \ref{fig:graph-iterations} the convergence
behavior of the proposed algorithm by plotting the average sum-MSE
versus the number of iterations for $N_{U}=6$ UEs, $N_{R}=2$ RRHs,
$M=4$ RRH antennas, $\tau=2$ pilot symbols and 10 dB SNR. From the
figure, we observe that the proposed algorithm converges within a
few iterations.

\begin{figure}
\centering\includegraphics[width=8.65cm,height=6.15cm]{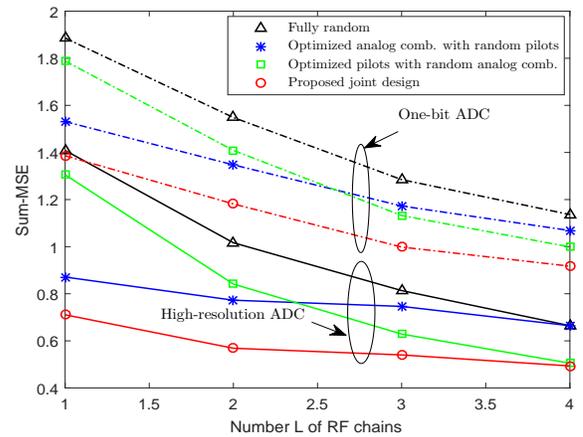}\caption{{\scriptsize{}{}\label{fig:graph-RFchains}}{\footnotesize{}{}Average
sum-MSE versus the number $L$ of RF chains ($N_{U}=6$, $N_{R}=2$,
$M=4$, $\tau=2$ and 10 dB SNR).}}
\end{figure}

In Fig. \ref{fig:graph-RFchains}, we investigate the impact of the
number $L$ of RF chains for the same configuration as in the previous
figure. A first observation is that optimizing analog combiner yields
larger performance gain for smaller values of $L$, where fewer signal
dimensions are available for channel estimation at the receiver. In
contrast, optimizing the pilots only provides more significant performance
gain for larger values of $L$. In this regime, the channel estimation
performance is dominated by the variance due to channel noise rather
than by the bias caused by a small number $L$ of RF chains. Joint
optimization allows both gains of optimizing pilots and analog combiners
to be harnessed. Finally, we note that, with one-bit ADCs, optimized
analog combining design offers performance gains even when $L=M$.
This is because the analog combiners can pre-process the received
signal in order to enable the one-bit ADCs to extract the most useful
information for channel estimation.

\begin{figure}
\centering\includegraphics[width=8.65cm,height=6.15cm]{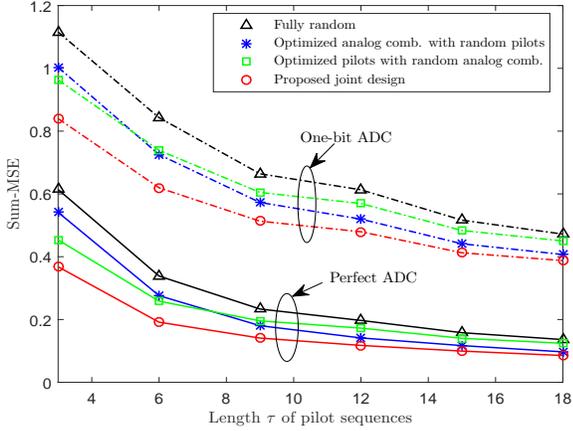}\caption{{\scriptsize{}{}\label{fig:graph-pilotLength}}{\footnotesize{}{}Average
sum-MSE versus the pilot length $\tau$ ($N_{U}=6$, $N_{R}=2$, $M=4$,
$L=3$ and 10 dB SNR).}}
\end{figure}

In Fig. \ref{fig:graph-pilotLength}, we plot the average sum-MSE
versus the pilot length $\tau$ for $N_{U}=6$ UEs, $N_{R}=2$ RRHs,
$M=4$ RRH antennas, $L=3$ RF chains and 10 dB SNR. It is observed
that the impact of pilot optimization is more significant in the regime
where $\tau$ is smaller, which calls for the use of well-designed
pilot signals.

\begin{figure}
\centering\includegraphics[width=8.65cm,height=6.15cm]{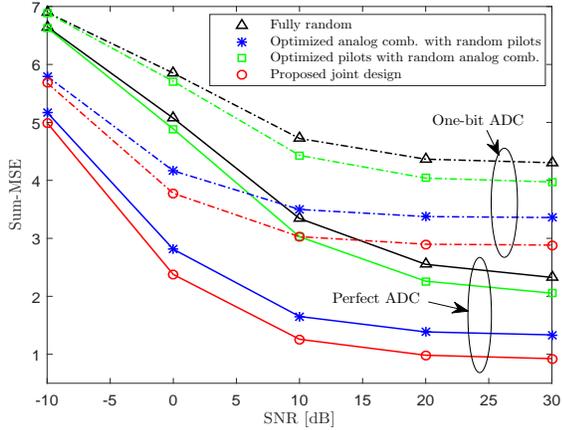}\caption{{\scriptsize{}{}\label{fig:graph-SNR}}{\footnotesize{}{}Average
sum-MSE versus the SNR ($N_{U}=6$, $N_{R}=2$, $M=10$, $L=2$ and
$\tau=3$).}}
\end{figure}

Lastly, Fig. \ref{fig:graph-SNR} plots the average sum-MSE versus
the SNR for $N_{U}=6$ UEs, $N_{R}=2$ RRHs, $M=10$ RRH antennas,
$L=2$ RF chain and $\tau=3$ pilot symbols. We note that the pilot
optimization has a negligible impact on the performance in the low
SNR regime, where the performance is limited by additive noise. In
contrast, the design of analog combiners provides relevant performance
gains even for low SNRs, since it can provide array beamforming gains
to increase the effective SNR at the combiners' output signals.

\section{Conclusions\label{sec:Conclusion} }

The joint design of the pilot signals
and analog combining matrices was addressed for a cell-free C-RAN system with the goal of minimizing
the sum-MSE metric of all the channel vectors in the presence of high-resolution or one-bit
ADCs. We observed that, with high-resolution ADCs and noiseless uplink channel, the analog combining matrix of each RRH can be separately optimized.
For the optimization with one-bit ADCs, we modeled the impact of ADC by leveraging Bussgang's
theorem, and proposed an iterative algorithm that alternately optimizes the pilots, analog combiners and digital filter matrices.
As a future work, we mention the analysis of the impact of fronthaul compression techniques for cell-free massive MIMO systems with
finite-capacity fronthaul links.


\begin{thebibliography}{15}
\bibitem{Quek-et-al}
T. Q. Quek, M. Peng, O. Simeone and W. Yu, \textit{Cloud Radio Access Networks: Principles, Technologies, and Applications}, Cambridge Univ. Press, Apr. 2017.
\bibitem{Sohrabi-Yu}
F. Sohrabi and W. Yu, "Hybrid digital and analog beamforming design for large-scale antenna arrays," \textit{IEEE Journ. Sel. Topics Sig. Proc.}, vol. 10, no. 3, pp. 501-513, Apr. 2016.
\bibitem{Ioushu-Eldar:SU}
S. S. Ioushua and Y. C. Eldar, "Hybrid analog-digital beamforming for massive MIMO systems," arXiv:1712.03485, Dec. 2017.
\bibitem{Ioushua-Eldar}
S. S. Ioushua and Y. C. Eldar, "Pilot contamination mitigation with reduced RF chains," arXiv:1801.05483, Jan. 2018.
\bibitem{Lee-et-al:TCOM}
C.-S. Lee and W.-H. Chung, "Hybrid RF-baseband precoding for cooperative multiuser massive MIMO systems with limited RF chains," \textit{IEEE Trans. Comm.}, vol. 65, no. 4, pp. 1575-1589, Apr. 2017.
\bibitem{Bussgang}
J. J. Busggang, "Crosscorrelation functions of amplitude-distorted Gaussian signals," \textit{Res. Lab. Electron.}, Massachusetts Inst. Technol., Cambridge, MA, USA, Tech. Rep. 216, 1952.
\bibitem{Jacobsson-et-al}
S. Jacobsson, G. Durisi, M. Coldrey, U. Gustavsson and C. Stuber, "Throughput analysis of massive MIMO uplink with low-resolution ADCs," \textit{IEEE Trans. Wireless Comm.}, vol. 16, no. 6, pp. 4038-4051, Jun. 2017.
\bibitem{Li-et-al}
Y. Li, C. Tao, G. Seco-Granados, A. Mezghani, A. L. Swindlehurst and L. Liu, "Channel estimation and performance analysis of one-bit massive MIMO systems," \textit{IEEE Trans. Sig. Proc.}, vol. 65, no. 15, pp. 4075-4089, Aug. 2017.
\bibitem{Kang-et-al}
J. Kang, O. Simeone, J. Kang and S. Shamai (Shitz), "Joint signal and channel state information compression for the backhaul of uplink network MIMO systems," \textit{IEEE Trans. Wireless Comm.}, vol. 13, no. 3, pp. 1555-1567, Mar. 2014.
\bibitem{JPark-et-al}
J. Park, S. Park, A. Yazdan and R. W. Heath, Jr., "Optimization of mixed-ADC multi-antenna systems for cloud-RAN deployments," \textit{IEEE Trans. Commun.}, vol. 65, no. 9, pp. 3962-3975, Sep. 2017.
\bibitem{Xu-Yin}
Y. Xu and W. Yin, "A block coordinate descent method for regularized multiconvex optimization with applications to nonnegative tensor factorization and completion," \textit{SIAM Journ. Imag. Scien.}, vol. 6, no. 3, pp. 1758-1789, 2013.
\bibitem{Scutari-et-al}
G. Scutari, F. Facchinei and L. Lampariello, "Parallel and distributed methods for constrained nonconvex optimization-Part I: Theory," \textit{IEEE Trans. Sig. Proc.}, vol. 65, no. 8, pp. 1929-1944, Apr. 2017.
\bibitem{Shiu-et-al}
D.-S. Shiu, G. J. Foschini, M. J. Gans and J. M. Kahn, "Fading correlation and its effect on the capacity of multielement antenna systems," \textit{IEEE Trans. Comm.}, vol. 48, no. 3, pp. 502-513, 2000.
\bibitem{Ioushua-Eldar:SPAWC}
S. S. Ioushua and Y. C. Eldar, "Pilot contamination mitigation with reduced RF chains," in \textit{Proc. IEEE Sig. Proc. Adv. Wireless Commun. (SPAWC 2017)}, Sapporo, Japan, Jul. 2017.
\end{thebibliography}
\end{document}